# Skyrmion lattice with a giant topological Hall effect in a frustrated triangular-lattice magnet

Takashi Kurumaji[1]*, Taro Nakajima[1], Max Hirschberger[1], Akiko Kikkawa[1], Yuichi Yamasaki[1,2,3], Hajime Sagayama[4], Hironori Nakao[4], Yasujiro Taguchi[1], Taka-hisa Arima[1,5], and Yoshinori Tokura[1,6]

[1] *RIKEN Center for Emergent Matter Science (CEMS), Wako 351-0198, Japan,*

[2] *Research and Services Division of Materials Data and Integrated System (MaDIS), National Institute for Materials Science (NIMS), Tsukuba 305-0047, Japan,*

[3] *PRESTO, Japan Science and Technology Agency (JST), Kawaguchi 332-0012, Japan,*

[4] *Institute of Materials Structure Science, High Energy Accelerator Research Organization, Tsukuba, Ibaraki 305-0801, Japan,*

[5] *Department of Advanced Materials Science, The University of Tokyo, Kashiwa 277-8561, Japan,*

[6] *Department of Applied Physics, The University of Tokyo, Tokyo 113-8656, Japan.*

*Corresponding author: takashi.kurumaji@riken.jp

**Abstract**

Geometrically frustrated magnets provide abundant opportunities for discovering complex spin textures, which sometimes yield unconventional electromagnetic responses in correlated electron systems. It is theoretically predicted that magnetic frustration may also promote a topologically nontrivial spin state, *i.e.*, magnetic skyrmions, which are nanometric spin vortices. Empirically, however, skyrmions are essentially concomitant with noncentrosymmetric lattice structures or interfacial-symmetry-breaking heterostructures. Here, we report the emergence of a Bloch-type skyrmion state in the frustrated centrosymmetric triangular-lattice magnet $Gd_2PdSi_3$. We identified the field-induced skyrmion phase via a giant topological Hall response, which is further corroborated by the observation of in-plane spin modulation probed by resonant x-ray scattering. Our results exemplify a new gold mine of magnetic frustration for producing topological spin textures endowed with emergent electrodynamics in centrosymmetric magnets.

In geometrically frustrated magnets, where competing interactions among localized spins cannot be simultaneously satisfied, conventional magnetic orders are suppressed to occur. Consequently, spins strongly fluctuate and form a disordered state known as spin liquid states (*1*), or occasionally find a route to various spin textures, including spin spiral orders or more complex noncoplanar orders (*2,3*). These spin states are mutually competing in energy, being highly susceptible to small perturbations showing a versatile magnetic phase diagram with respect to temperature, magnetic field, and pressure. One of the typical ways to distinguish a nontrivial feature of each emerging spin state is viewing it from the perspective of a geometrical correlation of spin vectors ($S_i$) on neighboring sites (*i, j, k*) in a lattice. For example, the vector spin chirality $S_i \times S_j$ describes the handedness of a spin spiral (*4*), and the scalar spin chirality $S_i \cdot (S_j \times S_k)$ is connected to time-reversal symmetry breaking (*5, 6*). These composite spin parameters couple with charge degrees of freedom in a correlated electron system to show unconventional electromagnetic responses (*7-10*). Exploration of novel spin textures via magnetic frustration has been one of the central issues of condensed matter physics.

A topological number of spin configurations, which remains intact under local deformation or weak fluctuations, offers another unique parameter to capture the characteristic nature of spin textures (*11*). This concept has recently attracted growing interest since the discovery of magnetic skyrmion states in chiral magnets (*12, 13*). The magnetic skyrmion is a vortex-like nanometric spin structure carrying a nontrivial topological number, where spin moments within a skyrmion wrap a sphere integer times (*14*). The particle nature of this spin texture with sensitivity to the electronic current and external electric/magnetic fields highlights magnetic skyrmions as potential information carries (*15*). Extensive studies have successfully identified skyrmion-hosting materials in the form of both bulk compounds (*16*) and multilayer thin-film structures (*17*). From those, one can establish an empirical design principle for skyrmions (*18,19*); that is to seek for the lack of inversion symmetry in the crystallographic lattice structure or at the interfaces. These asymmetries cause the relativistic Dzyaloshinskii-Moriya (DM) interaction (*20, 21*), which inherently prefers twisted spin configurations. More recently, this dogma has been challenged in theories (*22-24*) which propose to stabilize a skyrmion state in centrosymmetric lattices via magnetic frustration while experimental realization and observation of unconventional electronic responses have remained elusive.

In this paper, we demonstrate that the metallic magnet $Gd_2PdSi_3$, composed of a triangular-lattice network of Gd atoms (Fig. 1A) in the centrosymmetric hexagonal structure, hosts a skyrmion-lattice (SkL) state upon the application of a magnetic field (*H*) perpendicular to the triangular-lattice plane, which is robust down to the lowest temperature. The transition into the topological spin state is characterized by a prominent topological Hall response (*25, 26*) with a sharp contrast to the adjacent topologically-trivial magnetic phases. Using resonant x-ray scattering (RXS), we identify the long-range order of Gd spins modulated in the triangular lattice plane. The spin texture of the field-induced SkL phase is consistent with a triangular-lattice of Bloch-type skyrmions (Fig. 1B) on the basis of the analysis of the scattering intensity among different diffraction geometries.

Gd$_2$PdSi$_3$ belongs to a family of rare-earth intermetallics $R_2$PdSi$_3$ (*R*: rare earth) (*27*). Its crystal structure derives from the simple AlB$_2$-type structure with a triangular-lattice of *R* atoms sandwiching a nonmagnetic honeycomb-lattice layer composed of Pd and Si atoms (Fig. 1A). Due to the difference in atomic size, Si and Pd atoms order into a superstructure along both in- and out-of-plane directions (*28*), while the overall structure retains centrosymmetry (Fig. S1A). This excludes the DM interaction as a source of the skyrmion state. Instead, the Ruderman-Kittel-Kasuya-Yosida (RKKY) type interaction among the local 4*f* moments dominates as mediated by the conduction-electron's spin (*29-31*); RKKY interactions on the triangular network of 4*f* moments in $R_2$PdSi$_3$ are moderately frustrated (*32*) and show rich magnetic phases including modulated structures (*33*). Specifically, in Gd$_2$PdSi$_3$ metamagnetic transitions have been observed under a magnetic field perpendicular to the triangular lattice, which is accompanied by nonmonotonic variation of longitudinal and transverse transport properties (*34*). These features suggest strong coupling between conduction electrons and Gd spins, and promise the emergence of unconventional spin structures in the triangular-lattice network of Gd 4*f* moments.

First, we compare the magnetic phase diagram determined by the ac susceptibility ($\chi'$) for *H*||*c* in Gd$_2$PdSi$_3$ (Fig. 1C) with the contour mapping of the topological response of each phase probed by the topological Hall resistivity $\rho_{yx}^{\mathrm{T}}$ (Fig. 1D). Owing to the topological nature of skyrmions, they show characteristic emergent electrodynamic responses (*14*). In metallic materials, in particular, the scalar spin chirality of skyrmions acts like a fictitious magnetic field, which generates a transverse motion of electrons, that is the topological

Hall effect (THE) (*25, 26, 35*). In many cases, the transverse resistivity $\rho_{yx}$ is composed of three components;

$$\rho_{yx} = R_0 B + R_S M + \rho_{yx}^{\mathrm{T}}, \qquad (1)$$

where the first and the second terms are the normal and anomalous Hall resistivities proportional to the magnetic induction field *B* and the magnetization *M*, respectively, and the third term represents the topological component. Due to the feasibility of separating the first two terms with magnetization measurement, $\rho_{yx}^{\mathrm{T}}$ is sometimes well isolated, which makes a topological Hall component a good probe for a signature of the skyrmion or related topological spin state in various materials (*36*). As shown in Fig. 1C, peaks in $\chi'$ with respect to *H* (Fig. S2) define the phase boundaries for the three magnetic phases (IC-1, A, and IC-2) in addition to the paramagnetic (PM) state (*34*). In the *H-T* phase diagram, we overlay the contour plot of $\rho_{yx}^{\mathrm{T}}$ (Fig. 1D), which is deduced from the Hall resistivity measurements. The enhanced topological Hall signal appearing exclusively in the *A* phase region suggests that in Gd$_2$PdSi$_3$ the application of *H* induces topological phase transitions in the context of spin textures. The magnitude of the THE at the lowest temperature is as large as 2.6 µΩcm, which is one or two orders of magnitude larger than those of other skyrmion hosting materials such as MnSi (40 nΩcm under high pressure) (*25, 26, 35*) and FeGe (0.16 µΩcm in a thin film) (*37*). This must be partly due to a shorter wavelength of the spin modulation (~2.4 nm) (Figs. S5 and 3B), which squeezes the emergent magnetic flux of a skyrmion, in contrast to the relatively large size of skyrmions (10 ~ 100 nm) in typical noncentrosymmetric (chiral or polar) magnets (*36*).

To corroborate the observation of the THE in the *A* phase, we show a typical $\rho_{yx}$-*H* curve together with the *M* for *H*‖*c* at 2 K (Fig. 2A). A sharp positive enhancement of $\rho_{yx}$ is apparent in a region between two stepwise changes of *M* defining the first-order like transitions to and from the *A* phase. In the IC-2 phase and higher field region, on the contrary, $\rho_{yx}$ stays negative with nearly field-linear behavior, at least up to 140 kOe (Fig. S3A), where *M* is 13.7 $\mu_B$/f.u. approaching the saturation value expected for the value of local Gd moment. This nearly saturated phase in principle hosts a topologically trivial spin arrangement, allowing us to simulate the Hall response with the first two terms in Eq. (1). The black solid line in Fig. 2A shows the fit to the high-field data of $\rho_{yx}$. The fitting quality is excellent for all measured temperatures, which allows us to unambiguously isolate $\rho_{yx}^T$ from $\rho_{yx}$ as shown in Fig. 2B as well as Fig. 1C. We note that the quality of fitting is least affected by using a different formula, *e.g.*, assuming skew scattering type anomalous Hall effect (Fig. S4). Figure 2C show the evolution of peak in $\rho_{yx}^T$ with temperature, respectively. Continuous decrease of $\rho_{yx}^T$ towards zero around 20 K suggests that this response is affected by the magnitude of the molecular field from 4*f* moment on the conduction electron through the *f-d* coupling, consistent with the scalar spin chirality model for THE (*35*).

To further examine the nature of the SkL state in the $A$ phase, we present the Hall resistivity as a function of the angle between $H$ and the $c$ axis in the experimental configuration illustrated in the inset of Fig. 2D. At $\theta = 0°$ ($H\|c$) with $H = 9.9$ kOe in the $A$ phase, $\rho_{yx}$ starts from a large positive value. As $H$ rotates clockwise away from the $c$ axis, the value of $\rho_{yx}$ remains flat until an abrupt drop to near zero at around $\theta = 45°$. A hysteresis with the width of ~ 15° is observed between clockwise and counter clockwise rotation scans of $H$, pointing to the first-order nature of this $H$-direction sensitive phase transition. As frequently observed in thin film systems (*38, 39*), when the SkL are confined in a two-dimensional space, this topological spin texture survives only in an $H$ oriented nearly perpendicular to the lattice plane. Similar behavior may be expected for the present system composed of stacked triangular-lattice layers. The above observation provides a measure of the topological number for the spin texture where the topological Hall signal sharply transitions from finite to zero upon the destabilization of the SkL state. In contrast, at $H = 40$ kOe far above the upper critical field of the $A$ phase, a smooth evolution of $\rho_{yx}$ is observed with negligible hysteresis. This high field behavior follows $\cos\theta$ (black solid line in Fig. 2D) indicating that $M$ closely follows the rotating $H$ and that the projection of $M$ and $B$ to the $c$ axis produce the first two terms in Eq. (1) as dominant contributions to $\rho_{yx}$ outside the $A$ phase region.

Having identified the emergence of a topological electromagnetic response in the $A$ phase, we examined the Gd spin structure under $H$ along the $c$ axis, by the magnetic RXS in resonance with the Gd $L_2$ edge. We observed the magnetic modulation along in-plane directions represented by the reciprocal-space vector $Q_1 = (q, 0, 0)$ (and equivalent $Q_2 = (0, -q, 0)$ and $Q_3 = (q, -q, 0)$ as well), in the magnetically ordered phase (*40*). Here, $q$ (~ 0.14) is the magnetic modulation wavenumber. In Figs. 3A and 3B, we show $M$ together with $q$ at 5 K as a function of $H$ along the $c$ axis. To define the phase boundary for each phase, we show the difference between the $M$ for the $H$-increasing and decreasing scans, $\Delta M$ (Fig. 3A). In the IC-1 phase, $q$ is almost independent with $H$, and starts to gradually increase entering the *A* phase with additional increase in the IC-2 state. Despite the clear first-order nature for each transition (vertical gray lines), $q$ shows merely a weak kink at each phase boundary and changes only 4 % in total between 0 Oe in the IC-1 phase and 20 kOe in the IC-2 phase. The orientation of the $Q$ vectors with respect to the triangular lattice does not change across these metamagnetic transitions. This restricts the candidate spin textures for each phase to the spin modulations with one or several equivalent $Q$ vectors plus a component of homogeneous magnetization ($q = 0$) along the $c$ axis. This is consistent with the intermediate-field SkL state which is represented as the superposition of three spiral spin modulations with their magnetic modulation vectors lying in the triangular-lattice plane and pointing 120° away from each other.

Figure 3C shows the $H$ dependence of the scattering intensities for respective satellite peaks for the three $Q$ vectors measured around a Bragg spot (2, 2, 0) in the $H$-decreasing scan. Starting from the high-field IC-2 phase region (10 kOe < $H$ < 20 kOe), we observed that the intensity for one of the $Q$ vectors ($I_{Q2}$) is markedly weak compared with $I_{Q1}$ and $I_{Q3}$. This suggests that the IC-2 state is a single-$Q$ spin state, and that volume fraction of each $Q$-domain is not equivalent. This imbalance among the equivalent three single-$Q$ domains may be due to residual strains on the sample induced by shaping and attaching on the sample holder (*41*). We propose that possible magnetic structures for the IC-2 state are a fan-like structure or a transverse conical structure, which both lack a finite scalar spin chirality in accord with the absence of a topological contribution in $\rho_{yx}$. With decreasing $H$, the intensities for all the three $Q$ vectors show a step-wise increase upon entering the $A$ phase. These simultaneous increases of intensity for all $Q$ cannot be reconciled with a single-$Q$ model as discussed for the IC-2 phase, and rather points to the multiple-$Q$ nature of the $A$ phase. Further decreasing $H$, the intensity for each $Q$ vector is almost unchanged while a prominent peak in $\Delta M$ (Fig. 3A) suggests the first-order phase transition from the $A$ phase to the IC-1 phase. This behavior suggests that the IC-1 state may possibly be of triple-$Q$ nature as well but forms topologically different spin texture than that of the $A$ phase.

In order to clarify the triple-$Q$ states in the $A$ and the IC-1 phases, we further analyzed the spin structure by relative intensities of magnetic satellite peaks observed around different fundamental Bragg reflections from chemical lattice. In Fig. 3D, we summarize the normalized intensities for three satellite $Q$ vectors for each phase around the three Bragg spots, (0, 4, 0), (2, 2, 0), and (4, 0, 0), with different diffraction geometries (Fig. 3E). We present the simulated intensity for the spin structures with the proper screw type and the cycloidal type modulations in Fig. 3D, respectively (*40*). Here we emphasize that these intensities are calculated by assuming that modulation amplitudes for the three $Q$-vectors are equal to each other, *i.e.,* the triple-$Q$ state. The quantitative comparison between the calculated and observed intensities reveals that the magnetic structures in the IC-1 and *A* phases can be reproduced by the superposition of three proper screw helical modulations. This suggests that the spin configuration of the skyrmions in the *A* phase is Bloch type as depicted in Fig. 1B. Preference to the Bloch-type spin configuration in the present system rather than the Néel type or the antiskyrmion type ones is consistent with the energetics of skyrmions affected by the dipole-dipole interaction (*14, 42*), which may play a significant role in the present Gd compounds. The observed intensities in the IC-2 state, on the other hand, clearly deviates from the simulations due perhaps to the multidomain effect, being consistent with the single-$Q$ nature of this phase as discussed above. In the case for constructing candidates of the spin textures for the IC-1 phase, we note that there remains a degree of freedom for relative phase among three helical modulations (*40*). When the relative-phase ($\varphi_i$) for each $Q_i$-vector is 0 (mod $2\pi$), the triple-$Q$ state represents Bloch-type SkL state as exemplified by the *A* phase in the present case. In the case for another relative-phase configuration such as $\varphi_i = \pi/6$, it composes a triangular-lattice of merons and antimerons (see Fig. S6 in the supplemental materials) with no net scalar spin chirality at zero field; this is compatible with the observed features for the IC-1 state.

The emergence of a triple-$Q$ zero-field ground state (IC-1) may be an interesting difference from the conventional nonocentrosymmetric skyrmion-hosting systems, which typically show a single-$Q$ helical state as the zero-field state (*36*). We also found unconventional features beyond the conventional helical or conical state in the IC-1 phase in the Hall effect. As shown in Fig. 2B, $\rho_{yx}^{\text{T}}$ starts to gradually increase from zero field prior to a steep increase characterizing the transition to the skyrmion state. This indicates an *H*-induced scalar spin chirality, and thus a non-coplanar spin nature, even in the IC-1 state. We believe that due to the frustration in the present system, versatile long-range modulated-spin orders compete with each other and generate rich multiple-*Q* structures. This intricate competition is absent in conventional noncentrosymmetric skyrmion-hosting materials with DM interaction, which favors a twist of spin to form a spiral spin ground state. Further investigation in the present system and related compounds may offer a deep insight into the interplay between magnetic frustration and topological spin texture with emergent electromagnetic responses.

In conclusion, we have shown that a Bloch-type SkL state is stabilized under *H*||*c* in the triangular lattice magnet Gd$_2$PdSi$_3$. The short magnetic-modulation period (~ 2.4 nm) originating from the RKKY interaction squeezes the emergent magnetic flux of the noncoplanar spin vortex generating a giant topological Hall response in the SkL phase. In addition to the enhanced Hall effect, theories predict that the skyrmion in a centrosymmetric lattice shows unique properties such as the compatible formation of antiskyrmion with skyrmion (*22, 23*) and a helicity-dependent current response (*42, 43*); these predicted features are absent in noncentrosymmetric systems with innate chirality or polarity dependent on the lattice structure. The conduction-electron mediated competing magnetic (RKKY) interactions on a geometrically frustrated lattice will provide a novel

platform for emergent electrodynamics due to topological spin textures and cultivate a cross field between spin topology and magnetic frustration.

**Acknowledgements**

The authors thank X. Z. Yu for efforts on Lorentz transmission electron microscopy experiments, and K. S. Okada, S. Hayami, H. Ishizuka, Y. Motome, and N. Nagaosa for enlightening discussions. X-ray scattering measurements were performed under the approval of the Photon Factory Program Advisory Committee (Proposals No. 2015S2-007) at the Institute of Material Structure Science, High Energy Accelerator Research Organization (KEK). This research was supported in part by Grant-in-Aid for Scientific Research(S) No. 24224009, Grant-In-Aid for Young Scientists(B) No. 17K14351, Grant-in-Aid for Scientific Research No. 16H05990 from the Japan Society for the Promotion of Science (JSPS), and by PRESTO No. JPMJPR177A from Japan Science and Technology Agency (JST).

**Figure Captions**

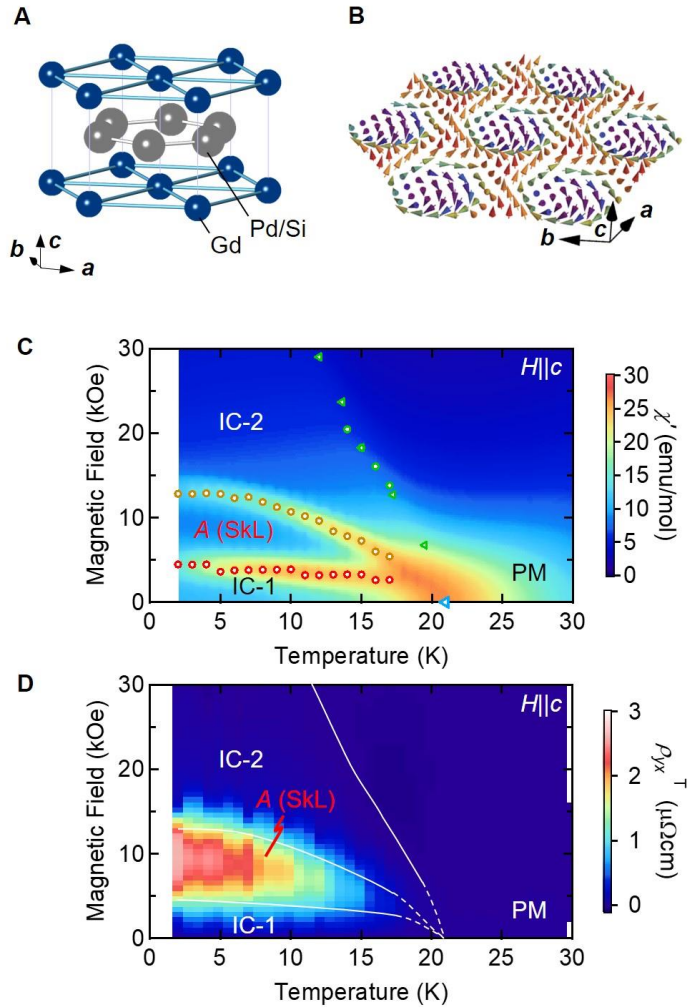

**Fig. 1. Phase diagram and topological Hall effect in Gd$_2$PdSi$_3$.** (**A**) The basic AlB$_2$-type crystal structure for Gd$_2$PdSi$_3$. (**B**) The schematic illustration of the spin texture in the skyrmion lattice (SkL) state. Each arrow indicates a spin at each Gd site. (**C**)-(**D**) The contour plot of (**C**) $\chi'$ and (**D**) that of $\rho_{yx}^{T}$ (see the text for the definition) for $H\|c$. "$A$" represents the SkL phase and "PM" the paramagnetic phase. The "IC-1" and "IC-2" phases

show incommensurate spin states in near-zero and high-field regions, respectively. Circular (triangular) symbols were determined by a peak or a kink in $\chi'$-$H$ ($\chi'$-$T$) scan (see Fig. S2).

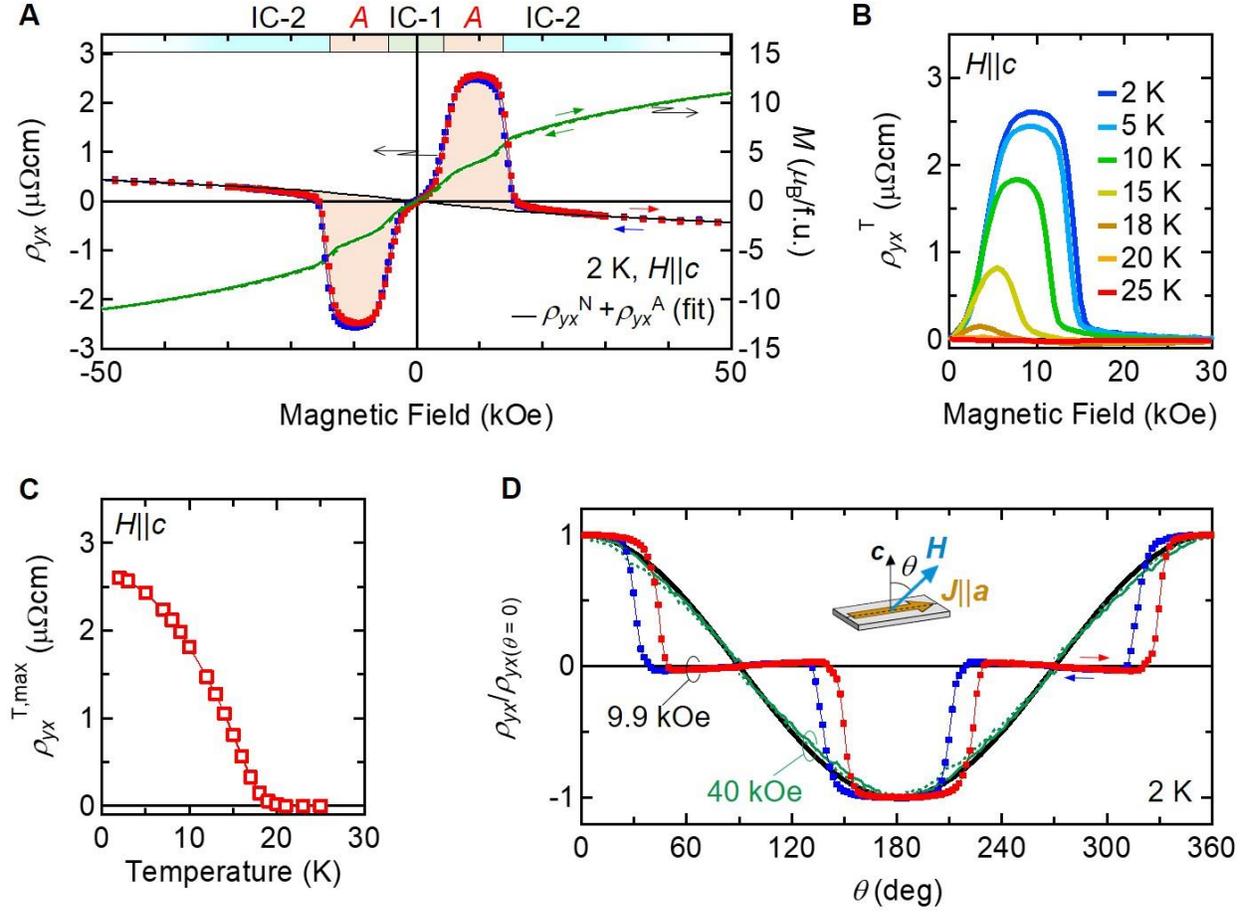

**Fig. 2. Temperature and angular dependence of the topological Hall effect in Gd$_2$PdSi$_3$.** (**A**) $H$ dependence of $\rho_{yx}$ and $M$ for $H\|c$ at 2 K. Red (blue) curve is for the $H$-increasing (-decreasing) scan. The black curve indicates the sum of the normal ($\rho_{yx}^{N}$) and the anomalous ($\rho_{yx}^{A}$) components of Hall resistivity. (**B**) $H$ dependence of topological Hall component $\rho_{yx}^{T}$ at various temperatures. (**C**) Temperature dependence of the maximum values of $\rho_{yx}^{T}$ ($\rho_{yx}^{T,max}$). (**D**) Normalized transverse resistivity at 2 K with $H$ rotating in the $ac$ plane. Red (blue) symbols and green solid (dashed) line are in a (counter-) clockwise

rotation. Inset defines the rotation angle $\theta$. The reference line $\cos\theta$ is shown by the black solid line.

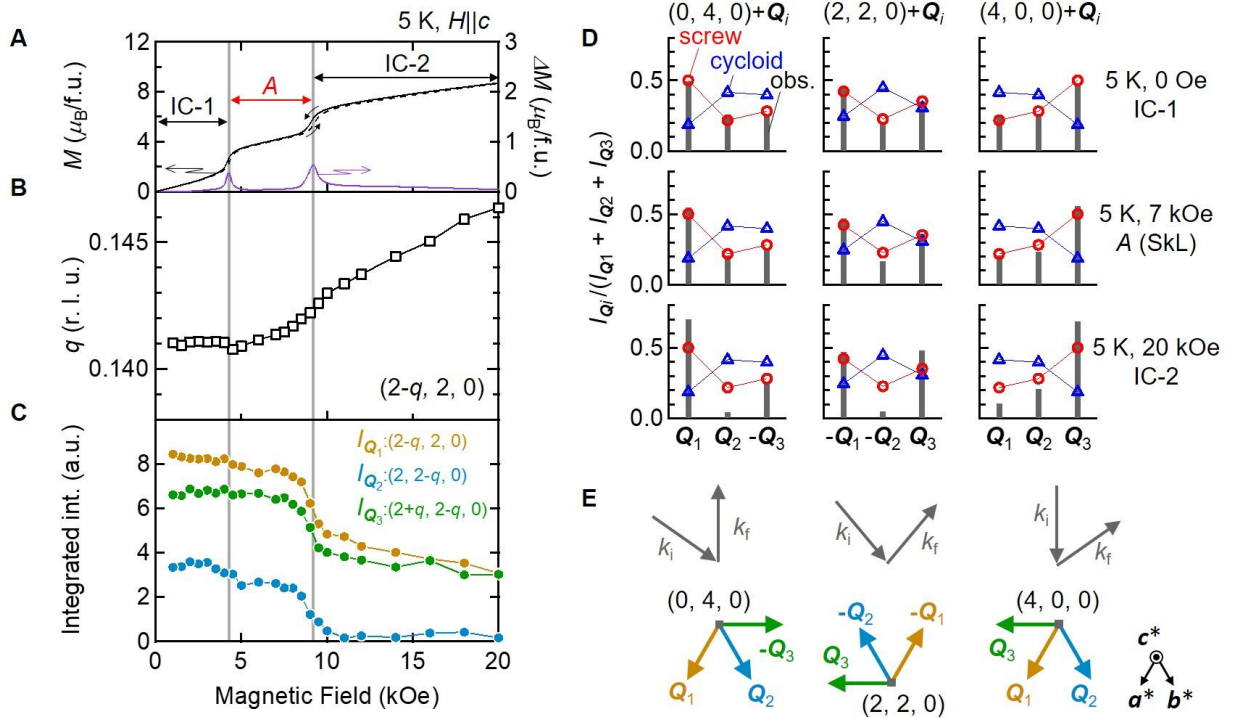

**Fig. 3. Analysis of spin textures by resonant x-ray scattering.** (**A**) $H$ dependence of $M$ in $H$-increasing (black dashed line) and decreasing (black solid line) processes and the difference of them ($\Delta M$, purple solid line). (**B**) $q$, and (**C**) integrated intensity for each magnetic satellite peak at $Q_i$ ($i = 1,2,3$) around the Bragg peak (2, 2, 0), measured at 5 K and in $H$-decreasing process. (**D**) Calculated and observed intensity for each satellite $Q_i$ at (0, 4, 0) (left), (2, 2, 0) (middle), and (4, 0, 0) (right), measured at 5 K with $H\|c$ of 0 Oe (top), 7 kOe (middle), and 20 kOe (bottom). (**E**) Respective diffraction geometries in terms of Bragg spots, incident ($k_i$) and scattered ($k_f$) beam vectors, and the modulation vectors $Q_i$ ($i = 1, 2, 3$) in the reciprocal space.

23